\documentclass[12pt]{article}
\usepackage{hyperref}
\usepackage{latexsym}
\usepackage{times}
\usepackage{amsmath}
\usepackage{graphicx}
\usepackage{xcolor}
\usepackage{placeins}
\usepackage[round]{natbib}

\newif\ifdraft \newif\ifblind
\draftfalse 
\blindfalse
\ifblind\newcommand{\blind}[1]{\colorbox{green}{\textcolor{green}{#1}}}\else\newcommand{\blind}[1]{#1}\fi

\newcommand{\keyw}[1]{\textcolor{red}{\emph{#1}}}

\newcommand{\Remark}[1]{\ifodd\value{page} \normalmarginpar
 \else \reversemarginpar \fi \marginpar{{\scriptsize #1}} }

\def\outdeg{\mathop{\rm outdeg}\nolimits}
\def\indeg{\mathop{\rm indeg}\nolimits}

\newcommand{\Graph}{\mathbf{G}}

\newcommand{\Edges}{{\cal E}}
\newcommand{\Nodes}{{\cal V}}

\newcommand{\clock}{\count254=\time \divide\count254 by 60
 \count255=\count254 \multiply\count255 by -60
 \advance\count255 by \time
 \ifnum\count254<10 0\fi\number\count254\,:\,%
 \ifnum\count255<10 0\fi\number\count255}

\title{Corrected overlap weight and clustering coefficient}
\ifblind\author{Author\\ Institution\\ e-mail}\else
\author{Vladimir Batagelj \\
        Institute of Mathematics, Physics and Mechanics,\\
         Department of Theoretical Computer Science,\\
        Jadranska 19, 1\,000 Ljubljana, Slovenia \\
        and \\
       University of Primorska, Andrej Marušič Institute,\\
       Muzejski trg 2, Koper, Slovenia \\   
       and \\
      National Research University Higher School of Economics,\\  Myasnitskaya, 20, 101000 Moscow, Russia    \\[3pt]
        e-mail: \texttt{vladimir.batagelj@uni-lj.si}\\ ORCID: 0000-0002-0240-9446}
\fi
\ifdraft\date{\today\ / \clock}\else\date{}\fi

\graphicspath{{./}{./pics/}{../../papers/2015/CRoNoS/london/slides/pics/}}

\begin{document}
\maketitle

\begin{abstract}
We discuss two well known network measures: the overlap weight of an edge and the
clustering coefficient of a node. For both of them it turns out that they are not
very useful for data analytic task to identify important elements (nodes or links) of a given network.
The reason for this is that they attain their largest values on maximal subgraphs of
relatively small size that are more probable to appear in a network than that of larger size.
We show how the definitions of these measures can be corrected in such a way that they give the expected
results. We illustrate the proposed corrected measures by applying them on the US Airports
network using the program Pajek.
\\[4pt]

\end{abstract}

\section{Introduction}

\subsection{Network element importance measures}

To identify important / interesting elements (nodes, links) in a network we often try to express
our intuition about their importantance using an appropriate measure (node index, link weight) following
the scheme\medskip

larger is the measure value of an element,  more important / interesting is this element.\medskip

\noindent
Too often, in analysis of networks, researchers uncritically pick some measure from the literature 
(degrees, closeness, betweenness, hubs and authorities, clustering coefficient, etc. \citep{WF,QSAR})
and apply it to their network.

In this paper we discuss two well known network local density measures: the overlap weight of an edge \citep{Oa} and 
the clustering coefficient of a node \citep{HL,WS}. 

For both of them it turns out that they are not
very useful for data analytic task to identify important elements of a given network.
The reason for this is that they attain their largest values on maximal subgraphs of
relatively small size -- they are more probable to appear in a network than that of larger size.
We show how their definitions can be corrected in such a way that they give the expected
results.  We illustrate the proposed corrected measures by applying them on the US Airports
network using the program Pajek. We will limit our attention to undirected simple graphs $\Graph =  (\Nodes,\Edges)$. 

Many similar indices and weights were proposed by graph drawing community for disentanglement in visualization of hairball networks \citep{edgemet,untangle,adaptive}.

When searching for important subnetworks in a given network we often assume a model that in the evolution of the network the increased activities in a part of the network create new nodes and edges in that part  increasing its local density. We expect from a \keyw{local density} measure $ld(x,\Graph)$ for an element (node/link) $x$ of network $\Graph$ the following properties:

\begin{itemize}
\item[\textbf{ld1.}] adding an edge, $e$, to the local neighborhood, $\Graph^{(1)}$, does not decrease the local density\\
 $ld(x,\Graph) \leq ld(x,\Graph \cup e)$.
\item[\textbf{ld2.}] normalization: \quad $0 \leq ld(x,\Graph) \leq 1$.
\item[\textbf{ld3.}] $ld(x,\Graph)$ can attain value 1, $ld(x,\Graph) = 1$, on the largest subnetwork of certain type in the network.
\end{itemize}

\section{Overlap weight}

\subsection{Overlap weight}

\begin{figure}[!h]
\centerline{\includegraphics[width=0.5\textwidth,viewport=90 65 560 452,clip=]{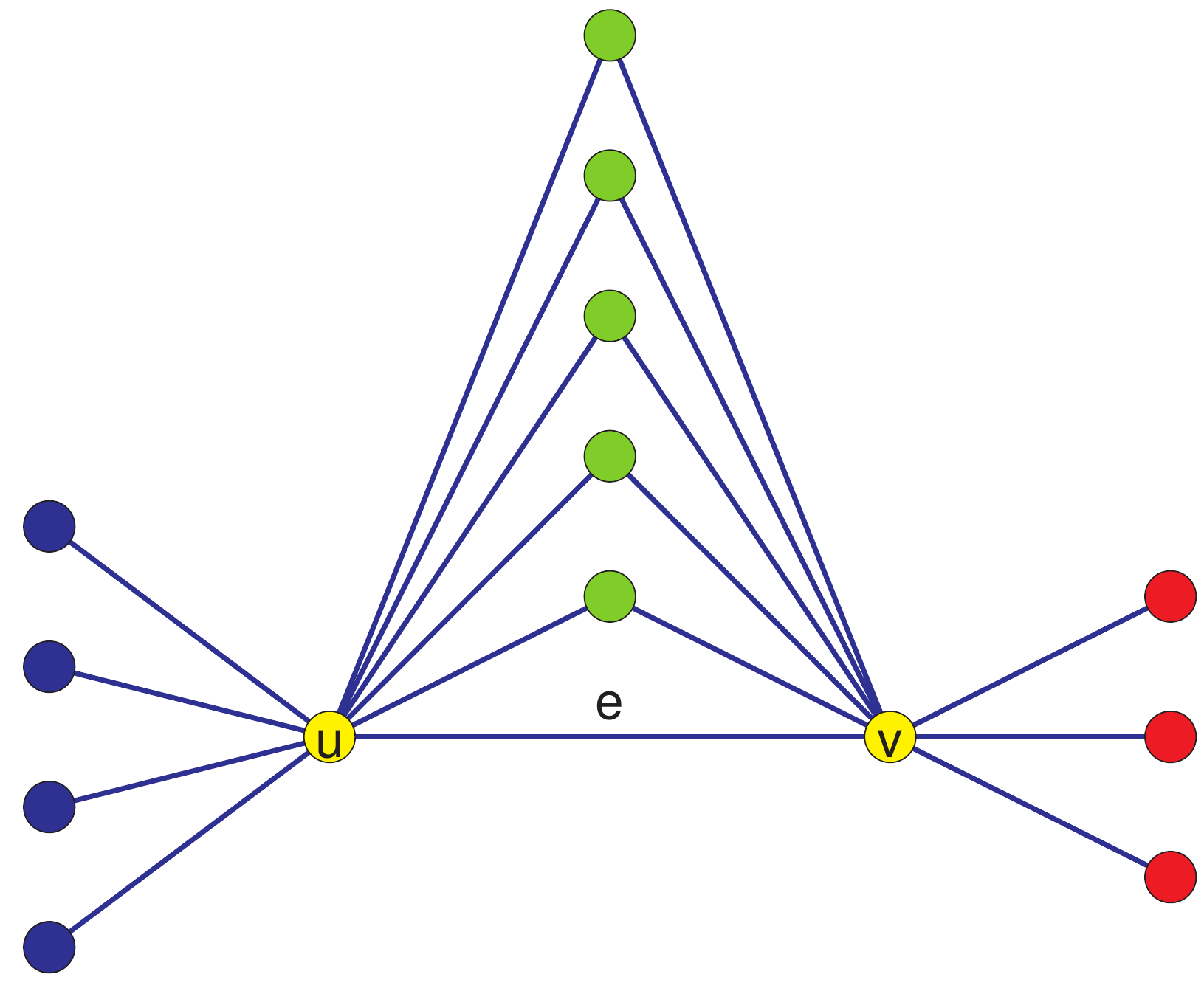}}
\caption{Neighbors of $e(u:v)$\label{euv}}
\end{figure}

A direct measure of the overlap  of an edge $e=(u:v) \in \Edges$  in an undirected simple graph $\Graph = (\Nodes,\Edges)$  is the number of common neighbors of its end nodes $u$ and $v$ (see Figure~\ref{euv}). It is equal to $t(e)$ --  the \keyw{number of triangles} (cycles of length 3) to which the edge $e$ belongs. The \keyw{edge neighbors subgraph} is labeled $T(\deg(u)-t(e)-1, t(e), \deg(v) - t(e)-1)$ -- the subgraph in Figure~\ref{euv} is labeled $T(4,5,3)$. There are two problems with this measure:
\begin{itemize}
\item it is not normalized (bounded to $[0,1]$);
\item it does not consider the `potentiality' of nodes $u$ and $v$ to form triangles -- there are
\[\min(\deg(u),\deg(v)) - 1 - t(e)\]
nodes in the smaller set of neighbors that are not in the other set of
neighbors.
\end{itemize}
Two simple normalizations are:
 \[ \frac{t(e)}{n-2}\qquad \mbox{ or } \qquad \frac{t(e)}{\mu}\]
where $n=|\Nodes|$ is the number of nodes, and  $\mu = \max_{e \in \Edges} t(e)$ is the maximum number of triangles on an edge in the graph $\Graph$.

The (topological) \keyw{overlap weight} of an edge $e=(u:v) \in \Edges$ considers also
 the degrees of edge's end nodes and is defined as
\[  o(e) = \frac{t(e)}{(\deg(u)-1)+(\deg(v)-1) - t(e)} \]
In the case $\deg(u)=\deg(v)=1$ we set $o(e) = 0$. It somehow resolves both problems.

The overlap weight is essentially a Jaccard similarity index \citep{wp}
\[ J(X,Y) = \frac{|X \cap Y|}{|X \cup Y|}\]
for $X = N(u) \setminus \{ v \}$ and $Y = N(v) \setminus \{ u \}$ where $N(z)$ is the set of neighbors of a node $z$.
In this case we have $|X \cap Y| = t(e)$ and
\[ |X \cup Y| = |X| + |Y| - |X \cap Y| = (\deg(u)-1)+(\deg(v)-1) - t(e) . \]
Note also that $h(X,Y) = 1- J(X,Y) =  \frac{|X \oplus Y|}{|X \cup Y|}$ is the normalized Hamming distance \citep{wp}. 
The operation $\oplus$ denotes
the symmetric difference $X \oplus Y = (X \cup Y) \setminus (X \cap Y)$.

Another normalized overlap measure is the \keyw{overlap index} \citep{wp}
\[ O(e) = O(X,Y) = \frac{|X \cap Y|}{\max(|X|,|Y|)} = \frac{t(e)}{\max(\deg(u),deg(v)) - 1}.\]
Both measures $J$ and $O$, applied to networks, have some nice properties. For example: a pair of nodes
$u$ and $v$ are structurally equivalent iff $J(X,Y) = O(X,Y) = 1$. Therefore the overlap weight measures the \keyw{substitutiability} of one edge's end node by the other.

\begin{figure}[tpb]
\centerline{\includegraphics[width=\textwidth,viewport=645 280 805 405,clip=]{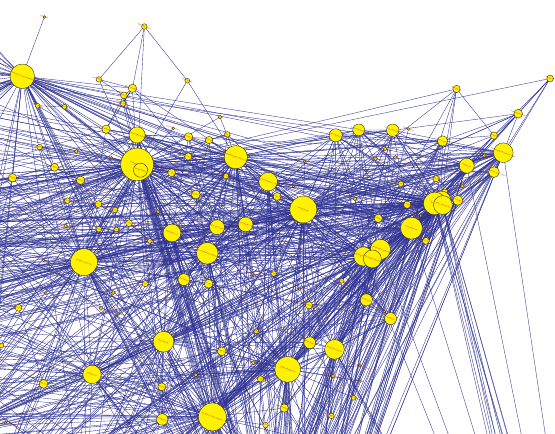}}
\caption{US Airports 1997 network, a North-East cut-out\label{USa}}
\end{figure}

Introducing two auxiliary quantities
\[ m(e) = \min(\deg(u),\deg(v)) - 1  \quad \mbox{and} \quad  M(e) = \max(\deg(u),\deg(v)) - 1  \]
we can rewrite the definiton of the overlap weight
\[  o(e) = \frac{t(e)}{m(e)+M(e) - t(e)},  \quad M(e) > 0 \]
and if $M(e)=0$ then $o(e)=0$.

For every edge $e \in \Edges$ it holds $ 0 \leq t(e) \leq m(e) \leq M(e)$.
Therefore
\[ m(e)+M(e)-t(e) \geq t(e)+t(e)-t(e) = t(e) \]
showing that $0 \leq o(e) \leq 1$.

The value $o(e)=1$ is attained exactly in the case when $M(e)=t(e)$; and the value $o(e)=0$ exactly when $t(e)=0$.

In simple directed graphs without loops different types of triangles exist over an arc $a(u,v)$. We can define
overlap weights for each type. For example: the \keyw{transitive overlap weight}
\[  o_t(a) = \frac{t_t(a)}{(\outdeg(u)-1)+(\indeg(v)-1) - t_t(a)} \]
and the \keyw{cyclic overlap weight}
\[  o_c(a) = \frac{t_c(a)}{\indeg(u)+\outdeg(v) - t_c(a)} \]
where $t_t(a)$ and $t_c(a)$ are the number of transitive / cyclic triangles containing the arc $a$. In this paper we will
limit our discussion to overlap weights in undirected graphs.

\subsection{US Airports links with the largest overlap weight}

Let us apply the overlap weight to the network of US Airports 1997 \citep{netData}. It consists of 332 airports
and 2126 edges among them. There is an edge linking a pair of airports iff in the year 1997 there was a flight company providing flights between those two airports.

The size of a circle representing an airport in Figure~\ref{USa} is proportional to its degree -- the number of
airports linked to it. The airports with the largest degree are: 

\begin{center}
\begin{tabular}{r|r}
airport & deg \\ \hline
Chicago O'hare Intl & 139 \\
Dallas/Fort Worth Intl & 118 \\
The William B Hartsfield Atlanta & 101 \\
Lambert-St Louis Intl & 94 \\
Pittsburgh Intl &  94 \\ \hline
\end{tabular} 
\end{center}

\begin{figure}
\centerline{\includegraphics[width=0.95\textwidth,viewport=9 143 680 582,clip=]{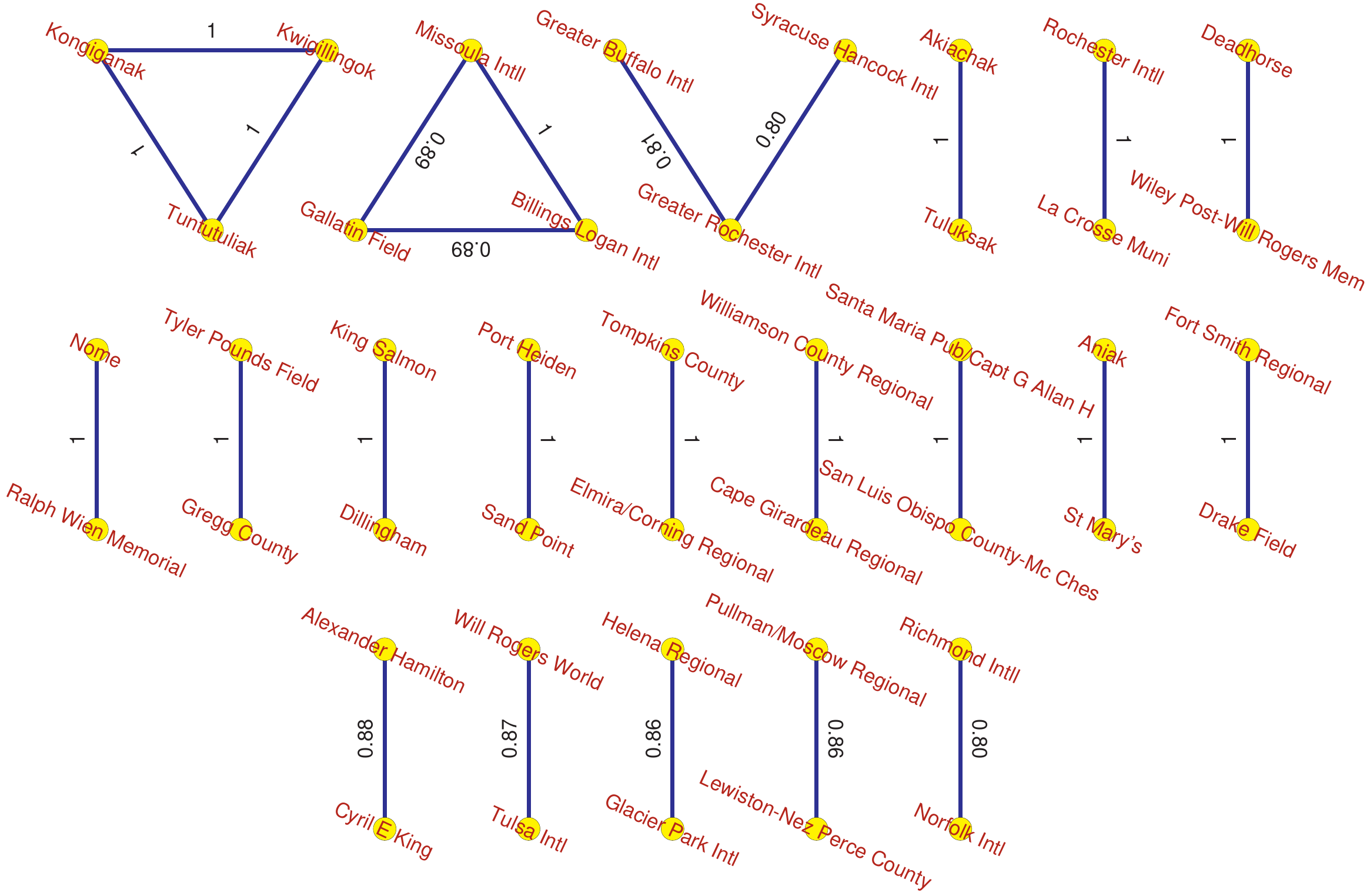}}
\caption{Edges with the largest overlap -- cut at 0.8\label{overCut}}
\end{figure}

For the overlap weight the edge cut at level 0.8 (a subnetwork of all edges with overlap weight at least 0.8) is
presented in Figure~\ref{overCut}. It consists of two triangles, a path of length 2, and 17 separate edges.

A tetrahedron (Kwigillingok, Kongiganak,Tuntutuliak, Bethel), see  Figure~\ref{tria}, gives the first
triangle in Figure~\ref{overCut} --
attached with the node Bethel to the rest of network.

\begin{figure}
\centerline{\includegraphics[width=0.5\textwidth,viewport=213 512 226 524,clip=]{UsAirMap.pdf}}
\caption{Zoom in \label{tria}}
\end{figure}

From this example we see that in real-life networks edges with the largest overlap weight tend to be edges with relatively small degrees in their end nodes ($o(e)=1$ implies $\deg(u) = \deg(v) = t(e)+1$) -- the overlap weight does not satisfy the condition \textbf{ld3}.
Because of this the overlap weight is not very useful for data analytic tasks in searching for important elements of a given network.
We would like to emphasize here that there are many applications in which overlap weight proves to be useful and
appropriate; we question only its appropriateness for determining the most overlaped edges.
We will try to improve the overlap weight definition to better suit the data analytic goals.

\subsection{Corrected overlap weight}

We define a \emph{corrected overlap weight} as
\[  o'(e) = \frac{t(e)}{\mu+M(e) - t(e)} \]

By the definiton of $\mu$ for every $e \in \Edges$ it holds $t(e) \leq \mu$. Since $M(e) - t(e) \geq 0$ also
$ \mu+M(e)-t(e)  \geq \mu $
and therefore \textbf{ld2}, $0 \leq o'(e) \leq 1$.  $o'(e)=0$ exactly when $t(e)=0$, and $o'(e)=1$ exactly when $\mu = M(e) = t(e)$. For \textbf{ld3}, the corresponding maximal edge neighbors subgraph contains $T(0,\mu,0)$. The end nodes of the edge $e$ are structurally equivalent.

To show that  \textbf{ld1} also holds let $\Graph^{(1)}(e)$ denote the edge neighbors subgraph of the edge $e$. Let $f$ be the edge added to  $\Graph^{(1)}(e)$ . We can assume that $\deg(u) \geq \deg(v)$, $e = (u:v)$. Therefore $M(e) = \deg(u) - 1$. We have to consider some cases: \\
\textbf{a.} $f \in \Edges(\Graph^{(1)}(e))$ : then $\Graph \cup f = \Graph$ and  $o'(e,\Graph \cup f) = o'(e,\Graph)$. \\
\textbf{b.} $f \notin \Edges(\Graph^{(1)}(e))$ : \\
\textbf{b1.} $f = (u:t)$ : then $t \in N(v)\setminus T(e) \setminus e$. It creates new triangle $(u,v,t)$. We have $t'(e) = t(e)+1$ and $M'(e) = M(e)+1$. We get
\[ o'(e,\Graph \cup f) = \frac{t'(e)}{\mu + M'(e) - t'(e)} =  \frac{t(e)+1}{\mu + M(e) - t(e)} > o'(e,\Graph) \]
\textbf{b2.}  $f = (v:t)$ : then $t \in N(u)\setminus T(e) \setminus e$. It creates new triangle $(u,v,t)$. We have $t'(e) = t(e)+1$ and $M'(e) = M(e)$. We get
\[ o'(e,\Graph \cup f) = \frac{t'(e)}{\mu + M'(e) - t'(e)} =  \frac{t(e)+1}{\mu + M(e) - t(e)-1} >  \frac{t(e)+1}{\mu + M(e) - t(e)} > o'(e,\Graph) \]
\textbf{b3.} $f = (t:w)$ and $t,w \in N(u) \cup N(v) \setminus \{u,v\}$ : No new triangle on $e$ is created. We have $t'(e) = t(e)$ and $M'(e) = M(e)$. Therefore  $o'(e,\Graph \cup f) = o'(e,\Graph)$.

The corrected overlap weight $o'$ is a kind of local density measure, but it is primarly a substitutiability measure. To get a better local density measure we have to consider besides triangles also quadrilaterals (4-cycles).

\subsection{US Airports 1997 links with the largest corrected overlap weight}

\begin{figure}
\centerline{\includegraphics[width=0.95\textwidth,viewport=14 185 736 630,clip=]{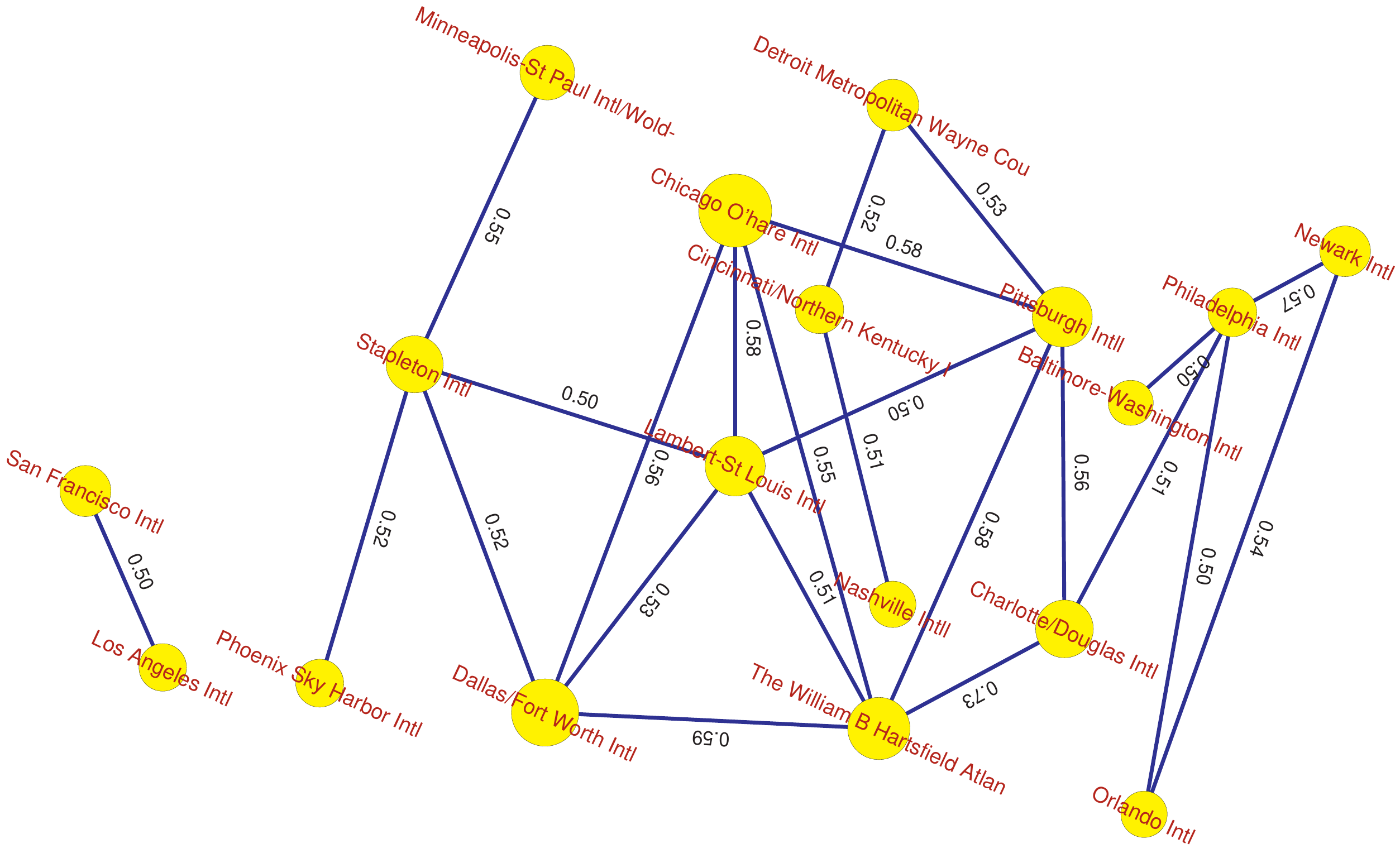}}
\caption{US Airports 1997 links  with the largest corrected overlap weight, cut at 0.5 \label{coverCut}}
\end{figure}

For the US Airports 1997 network we get $\mu = 80$. For the corrected overlap weight the edge cut at level 0.5 is
presented in Figure~\ref{coverCut}.
Six links with the largest triangular weights are given in Table~\ref{six}.

\begin{figure}
\centerline{\includegraphics[width=0.95\textwidth,viewport=15 25 883 618,clip=]{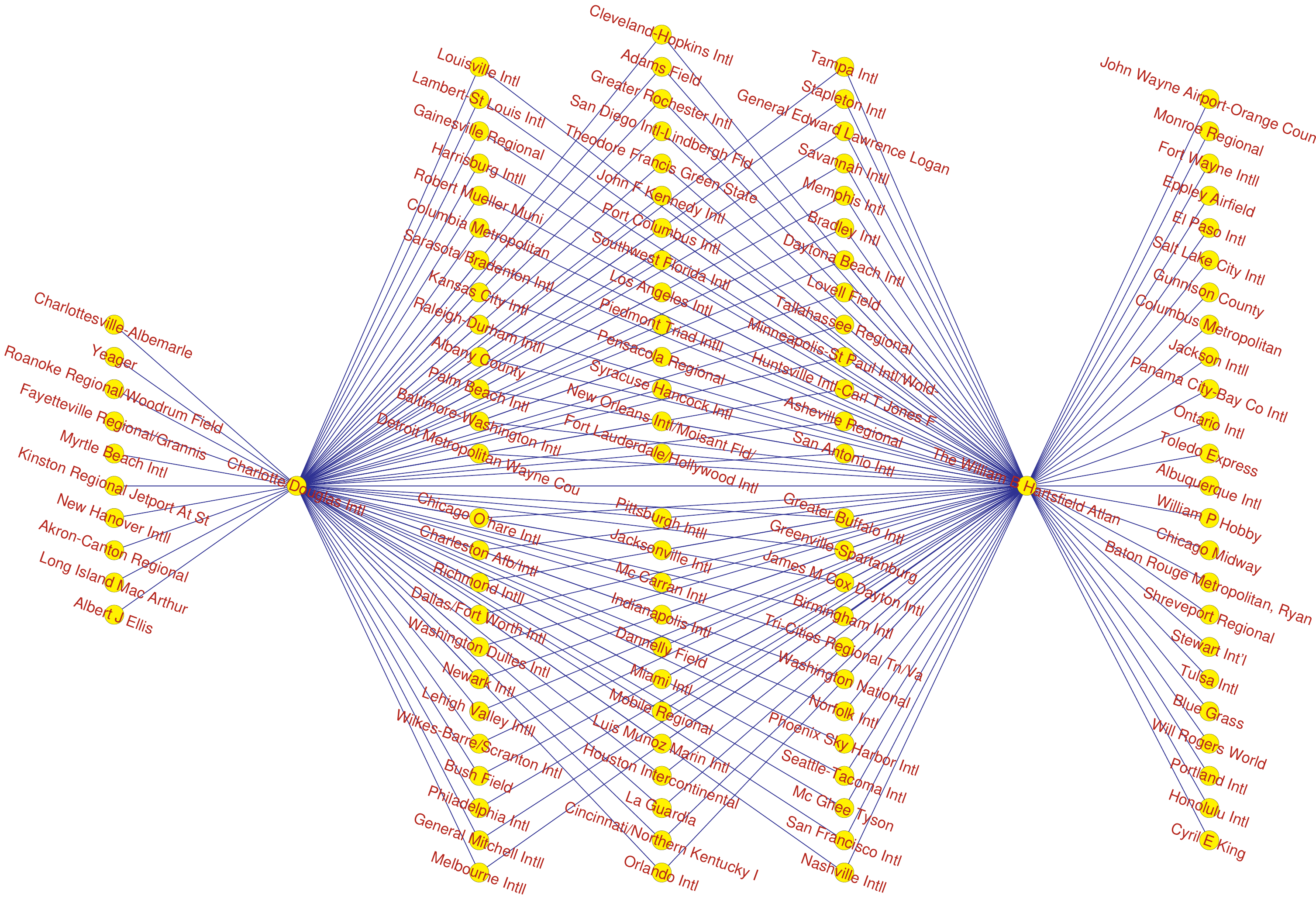}}
\caption{US Airports links $o'($WB Hartsfield Atlanta, Charlotte/Douglas Intl$) = 0.7308$ \label{link}}
\end{figure}

\begin{table}[h]
\caption{Largest triangular weights in US Airports 1997 network\label{six}}
\begin{tabular}{ll|rrrl|}
     $u$    &             $v$  &       $t(e)$ &  $d(u)$  &  $d(v)$  &   $o'(e)$   \\ \hline
Chicago O'hare Intl     &         Pittsburgh Intll   &  80 & 139 & 94  &  0.57971  \\
Chicago O'hare Intl     &         Lambert-St Louis Intl  &  80  & 139 & 94  &  0.57971 \\
Chicago O'hare Intl   &  Dallas/Fort Worth Intl     &      78  & 118 &139 &   0.55714  \\
Chicago O'hare Intl   &  The W B Hartsfield Atlanta &  77 & 101 & 139 &   0.54610 \\ 
The W B Hartsfield Atlanta &  Charlotte/Douglas Intl &  76 & 101 &  87 &   0.73077  \\
The W B Hartsfield Atlanta &  Dallas/Fort Worth Intl &  73 & 101 & 118 &   0.58871 \\
\hline
\end{tabular}
\end{table}

In Figure~\ref{link} all the neighbors of end nodes WB Hartsfield Atlanta and Charlotte/Douglas Intl
of the link with the largest corrected overlap weight value are presented. They have 76 common (triangular)
neighbors. The node  WB Hartsfield Atlanta has 11 and the node Charlotte/Douglas Intl has 25 additional
neighbors. Note  (see Table~\ref{six}) that there are some links with higher triangular weight, but also
with much higher number of additional neighbors -- therefore with smaller corrected overlap weights.

\subsection{Comparisons}

\begin{figure}
\centerline{\includegraphics[width=0.7\textwidth,viewport=0 5 390 360,clip=]{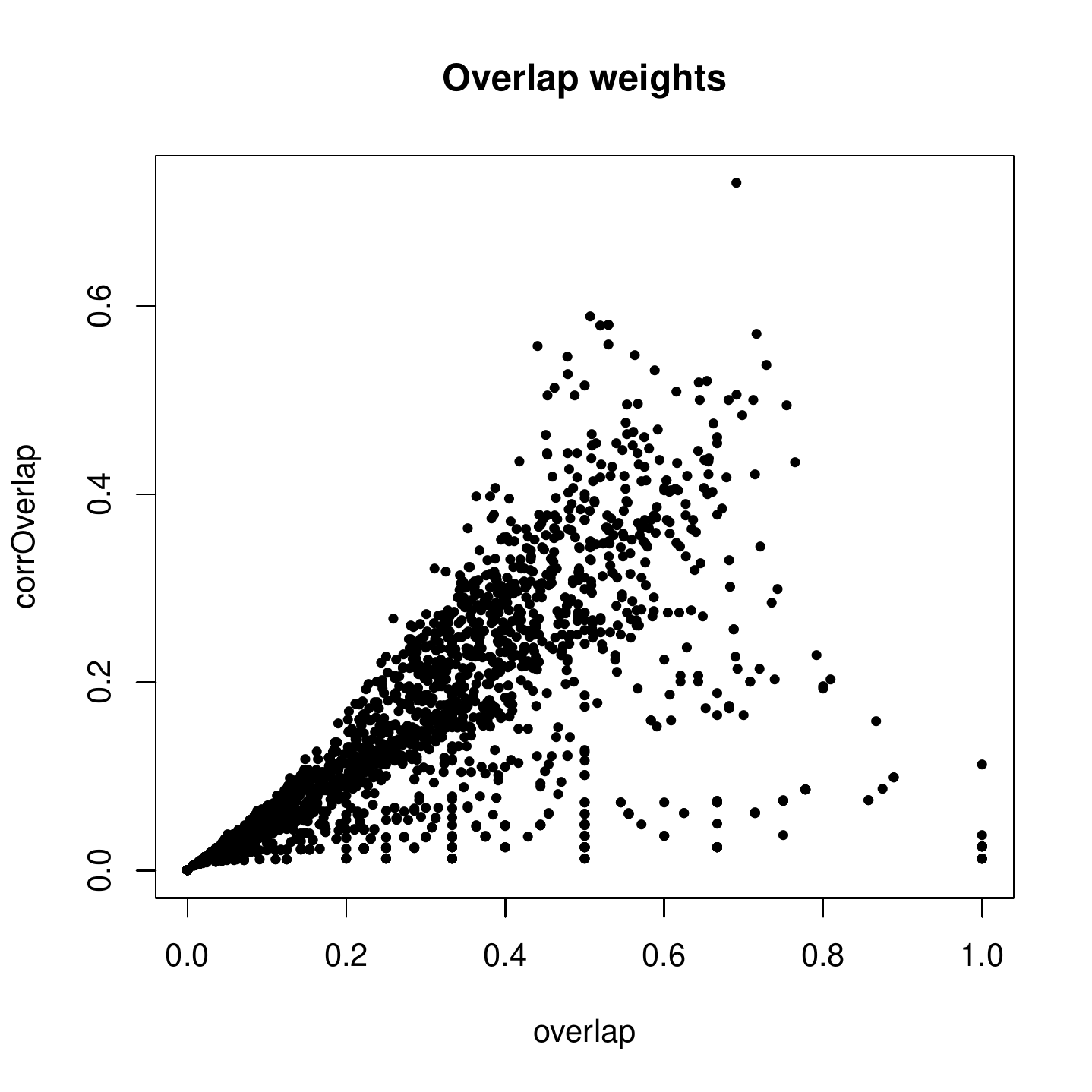}}
\caption{Comparison (overlap, corrected overlap) \label{oco}}
\end{figure}

In Figure~\ref{oco} the set $\{ (o(e), o'(e)) : e \in \Edges \}$  is displayed for the US Airports 1997 network.
For most edges it holds $ o'(e) \leq  o(e)$. It is easy to see that  $ o(e) <  o'(e) \Leftrightarrow \mu < m(e)$.
Edges with the  overlap value $o(e) > 0.8$ have the corrected overlap weight $o'(e) < 0.2$.

\begin{figure}
\centerline{\includegraphics[width=0.47\textwidth,viewport=0 5 390 360,clip=]{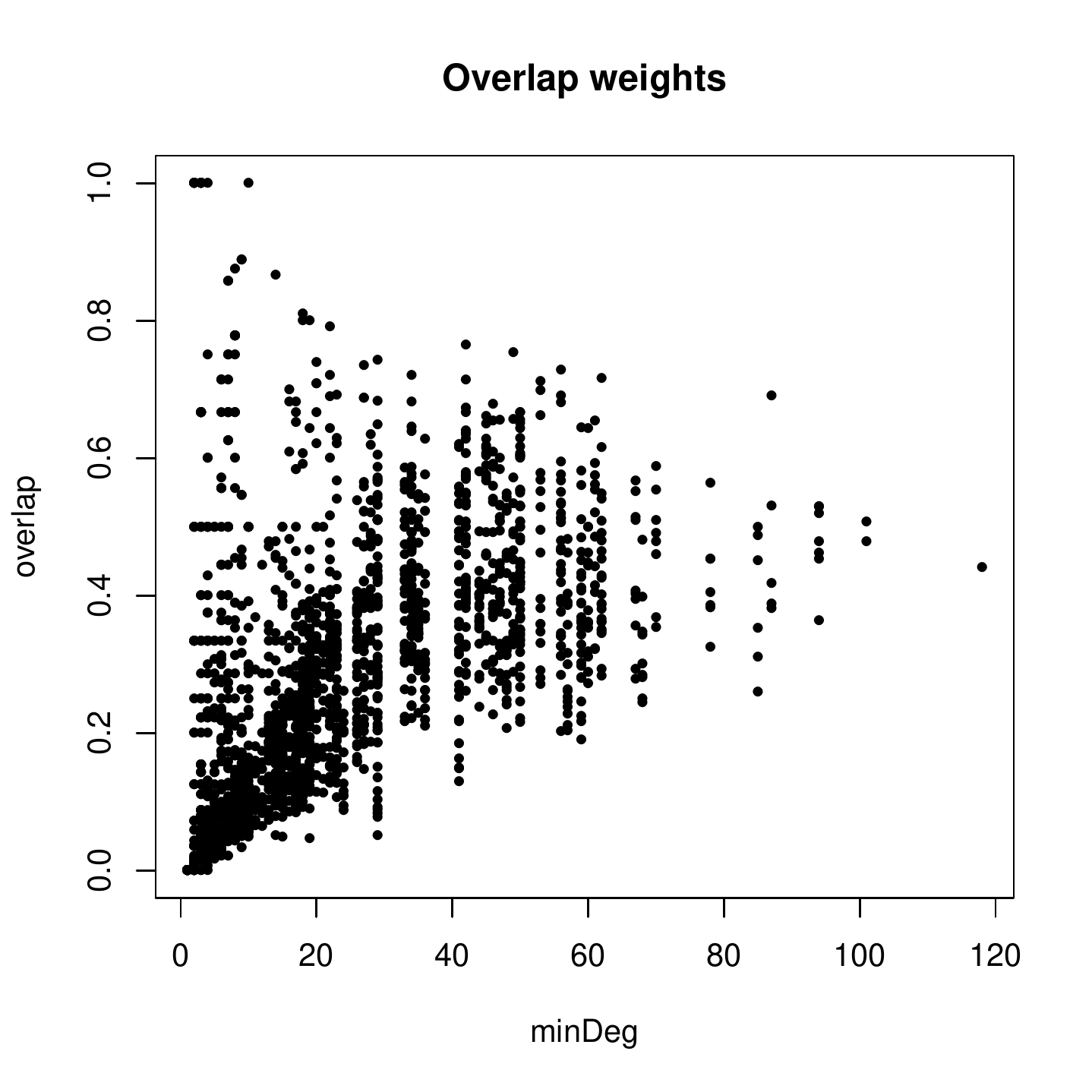}\qquad\includegraphics[width=0.47\textwidth,viewport=0 5 390 360,clip=]{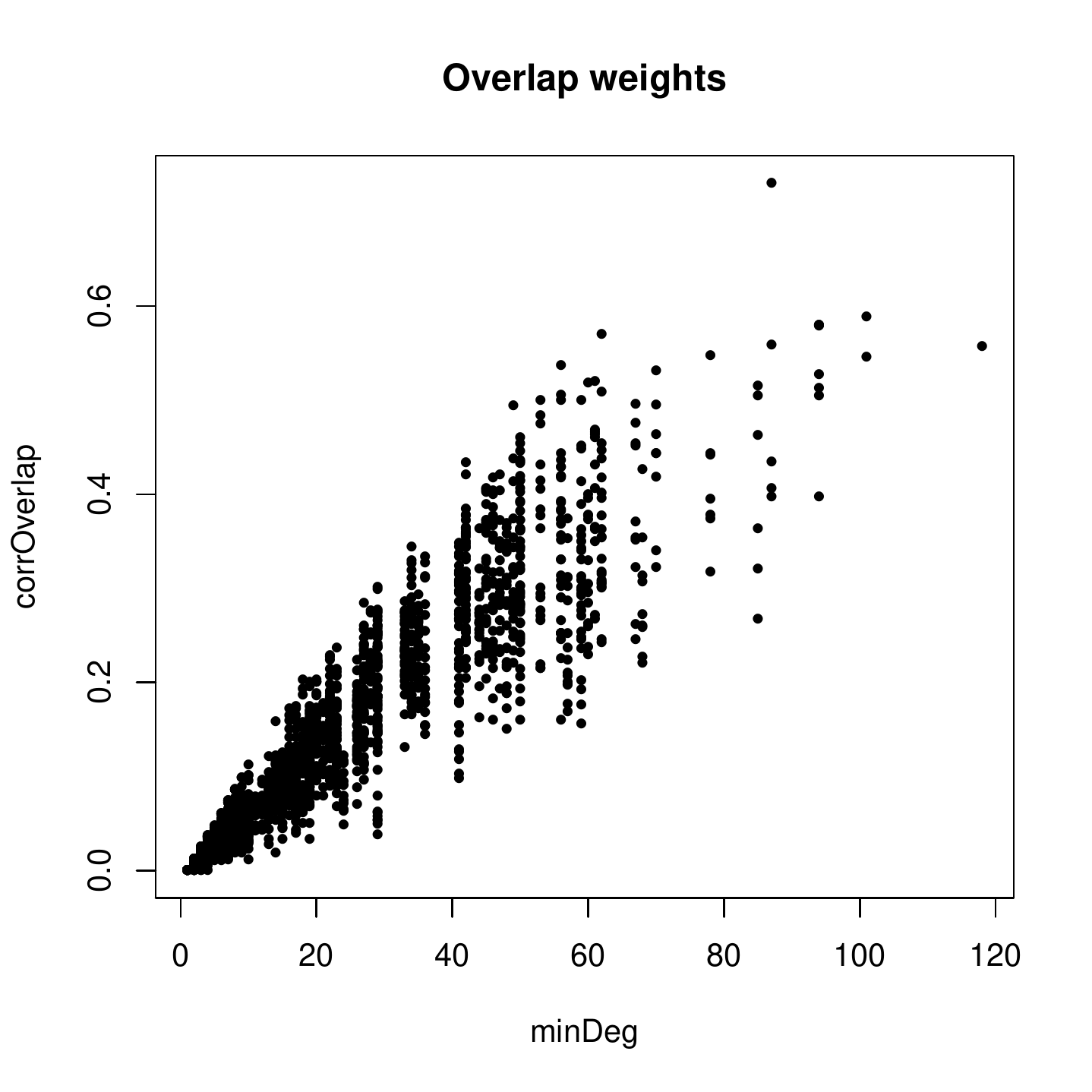}}
\caption{Comparison  -- minDeg$(e)$  \label{ocm}}
\end{figure}

In Figure~\ref{ocm} the sets $\{ (m(e), o(e)) : e \in \Edges \}$ and $\{ (m(e), o'(e)) : e \in \Edges \}$  are displayed for the US Airports 1997 network.
With increasing of $m(e)$ the corresponding overlap weight $o(e)$ is decreasing; and the corresponding corrected
overlap weight $o'(e)$ is also increasing.

\begin{figure}
\centerline{\includegraphics[width=0.47\textwidth,viewport=0 5 390 360,clip=]{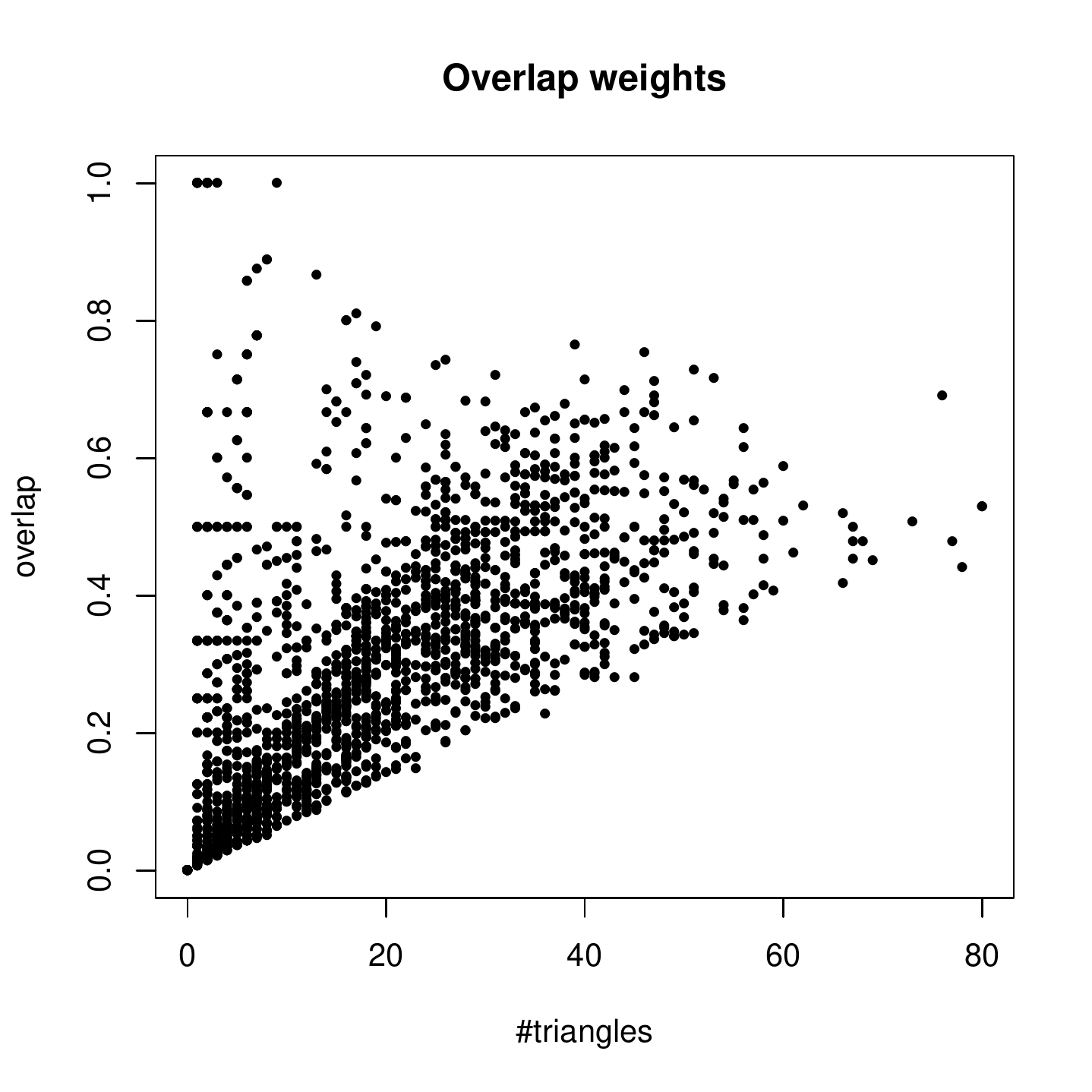}\qquad\includegraphics[width=0.47\textwidth,viewport=0 5 390 360,clip=]{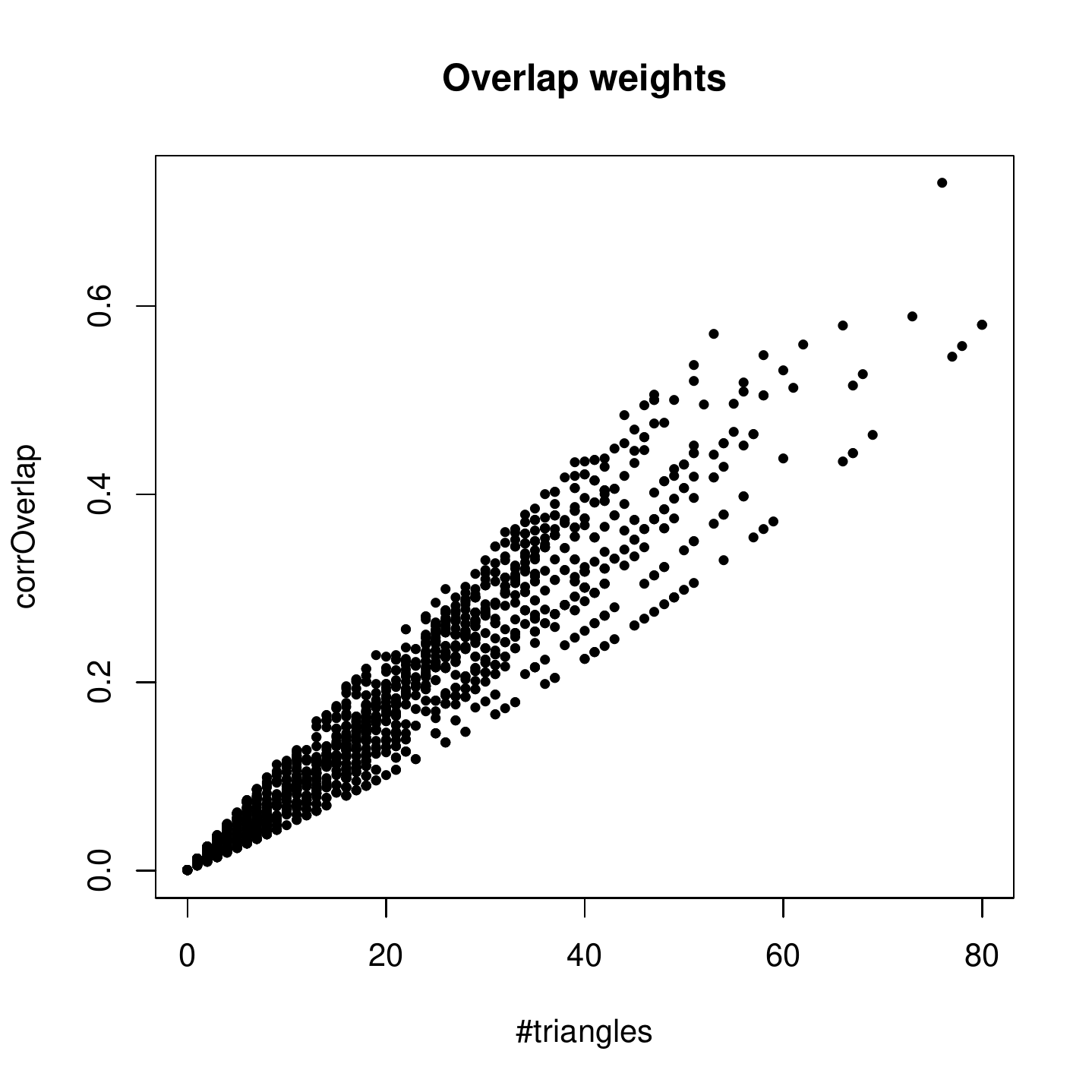}}
\caption{Comparison -- \# of triangles  \label{oct}}
\end{figure}

We can observe similar tendencies if we compare both weights with respect to the number of triangles $t(e)$ (see
Figure~\ref{oct}).

\FloatBarrier

\section{Clustering coefficient}

\subsection{Clustering coefficient}

For a node $u \in \Nodes$  in an undirected simple graph $\Graph = (\Nodes, \Edges)$ its (local) \keyw{clustering coefficient} \citep{wp}  is measuring a local density in the node $u$ and is defined as a proportion of the number of existing edges between $u$'s neighbors to the number of all possible  edges between $u$'s neighbors
\[ cc(u) = \frac{|\Edges(N(u))|}{|\Edges(K_{\deg(u)})|} = \frac{2\cdot  E(u)}{\deg(u)\cdot(\deg(u)-1)}, \quad \deg(u)>1 \]
where  $E(u) = |\Edges(N(u))|$. If $\deg(u) \leq 1$ then $cc(u) = 0$.\medskip

It is easy to see that
\[ E(u) = \frac{1}{2} \sum_{e \in S(u)} t(e)\]
where $S(u)=\{ e(u:v) : e \in \Edges\}$ is the star in node $u$. \medskip

It holds $0 \leq cc(u) \leq 1$; $cc(u) = 1$ exactly when $\Edges(N(u))$  is isomorphic to $K_{\deg(u)}$ -- a complete graph on $\deg(u)$ nodes. Therefore it seems that the clustering coefficient could be used to identify nodes with the densest neighborhoods.

The notion of clustering coefficient can be extended also to simple directed graphs (with loops).

\subsection{US Airports  with the largest clustering coefficient}

Let us apply also the clustering coefficient to the US Airports 1997 network.

\begin{table}
\caption{US Airports 1997  with  clustering coefficient $= 1$\label{CCmax}}
\begin{center}
\begin{tabular}{rrl|rrl|}
 $n$  &  $\deg$ &  airport		   &    $n$  &  $\deg$ &  airport	\\ \hline
    1  &   7 &  Lehigh Valley Intll	   &       8  &   4 &  Gunnison County		\\
    2  &   5 &  Evansville Regional	   &       9  &   4 &  Aspen-Pitkin Co/Sardy Field  \\
    3  &   5 &  Stewart Int'l		   &      10  &   4 &  Hector Intll		\\
    4  &   5 &  Rio Grande Valley Intl	   &      11  &   4 &  Burlington Regional	 \\
    5  &   5 &  Tallahassee Regional	   &      12  &   4 &  Rafael Hernandez		   \\
    6  &   4 &  Myrtle Beach Intl	   &      13  &   4 &  Wilkes-Barre/Scranton Intl \\
    7  &   4 &  Bishop Intll		   &      14  &   4 &  Toledo Express \\
 \hline
\end{tabular}
\end{center}
\end{table}

In Table~\ref{CCmax} airports with the clustering coefficient equal to 1 and the degree at least 4 are listed.
There are 28 additional such airports with a degree 3, and 38 with a degree 2.

Again we see that the clustering coefficient attains its largest value in nodes with relatively small degree. The probability that we get a complete subgraph on $N(u)$ is decreasing very fast with increasing of  $\deg(u)$. The  clustering coefficient does not satisfy the condition \textbf{ld3}.

\subsection{Corrected clustering coefficient}

To get a corrected version of the clustering coefficient we proposed \blind{in Pajek \citep{ESNA3}} to replace $\deg(u)$ in the denominator with $\Delta = \max_{v \in \Nodes} \deg(v)$. In this paper we propose another solution -- we replace $\deg(u)-1$ with $\mu$:
\[ cc'(u)  = \frac{2\cdot  E(u)}{\mu \cdot \deg(u)}, \quad \deg(u) > 0 \]
If $\deg(u)=0$ then $cc'(u)=0$.
Note that, if $\Delta > 0$ then $\mu < \Delta$.

To verify the property \textbf{ld1} we add to $\Graph(u)$ a new edge $f$ with its end nodes in $\Graph(u)$ . Then $E'(u) = E(u)+1$ and $\deg'(u) = \deg(u)$. Therefore
\[ cc'(u,\Graph \cup f)  = \frac{2\cdot  E'(u)}{\mu \cdot \deg'(u)} = \frac{2\cdot ( E(u)+1)}{\mu \cdot \deg(u)} > cc'(u,\Graph)\]
To show the property \textbf{ld2}, $0 \leq cc'(u) \leq 1$, we have to consider two cases:
\begin{itemize}
\item[\textbf{a.}] $\deg(u)\geq\mu$: then for $v \in N(u)$ we have $\deg_{N(u)}(v) \leq \mu$ and therefore
\[ 2\cdot E(u) =\sum_{v \in N(u)} \deg_{N(u)}(v) \leq \sum_{v \in N(u)} \mu = \mu \cdot \deg(u) \]
\item[\textbf{b.}] $\deg(u) < \mu$: then $\deg(u)-1 \leq \mu$ and therefore
\[ 2\cdot E(u) \leq \deg(u)\cdot(\deg(u)-1) \leq \mu \cdot \deg(u) \]
\end{itemize}
For the property \textbf{ld3}, the value $cc'(u)=1$ is attained in the case \textbf{a} on a $\mu$-core, and in the case \textbf{b} on $K_{\mu+1}$.

\subsection{US Airports nodes with the largest corrected clustering coefficient}

\begin{table}
\caption{US Airports 1997  with the largest corrected clustering coefficient \label{Lccc}}

 \begin{center}
\begin{tabular}{rrrl}
Rank &  Value &  deg & Id				\\ \hline
   1 & 0.3739 & 45 & Cleveland-Hopkins Intl          \\
   2 & 0.3700 & 50 & General Edward Lawrence Logan   \\
   3 & 0.3688 & 56 & Orlando Intl                    \\
   4 & 0.3595 & 42 & Tampa Intl                      \\
   5 & 0.3488 & 61 & Cincinnati/Northern Kentucky Intl  \\
   6 & 0.3457 & 70 & Detroit Metropolitan Wayne County  \\
   7 & 0.3455 & 67 & Newark Intl                     \\
   8 & 0.3429 & 53 & Baltimore-Washington Intl   \\
   9 & 0.3415 & 47 & Miami Intl                      \\
  10 & 0.3405 & 42 & Washington National             \\
  11 & 0.3379 & 56 & Nashville Intll                 \\
  12 & 0.3359 & 46 & John F Kennedy Intl             \\
  13 & 0.3347 & 62 & Philadelphia Intl               \\
  14 & 0.3335 & 41 & Indianapolis Intl               \\
  15 & 0.3335 & 50 & La Guardia                    \\ \hline
\end{tabular}
\end{center}
\end{table}

\begin{figure}
\centerline{\includegraphics[width=0.95\textwidth,viewport=0 0 520 520,clip=]{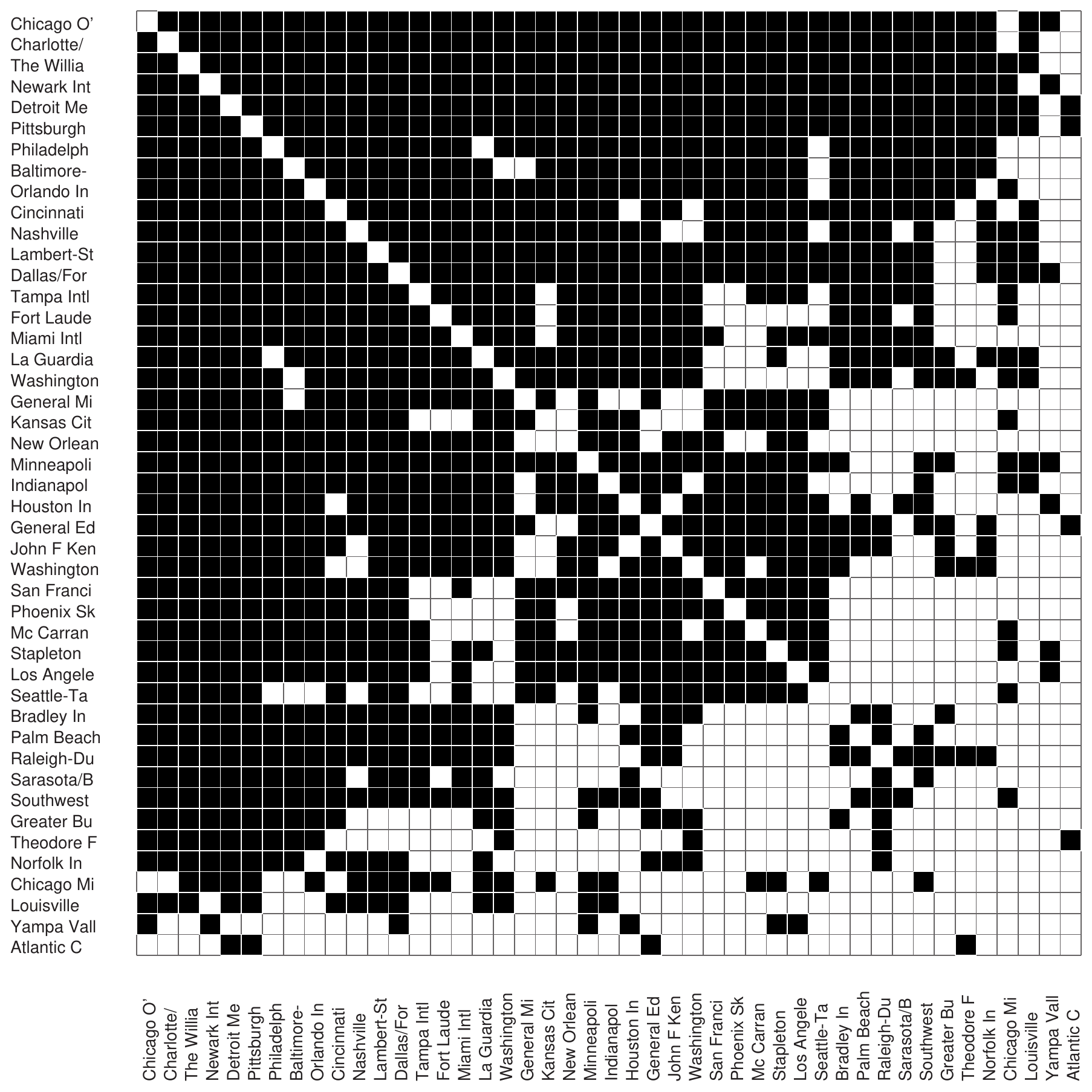}}
\caption{Links among Cleveland-Hopkins Intl neighbors\label{CH}}
\end{figure}

In Table~\ref{Lccc} US Airports with the largest corrected clustering coefficient are listed. The largest value 
0.3739 is attained for Cleveland-Hopkins Intl airport. In Figure~\ref{CH} the adjacency matrix of a subnetwork
on its 45 neighbors is presented. The subnetwork is relatively complete. A small value of corrected clustering coefficient 
is due to relatively small $\deg = 45$ with respect to $\mu = 80$.
 
\subsection{Comparisons}

\begin{figure}
\centerline{\includegraphics[width=0.47\textwidth,viewport=0 5 390 360,clip=]{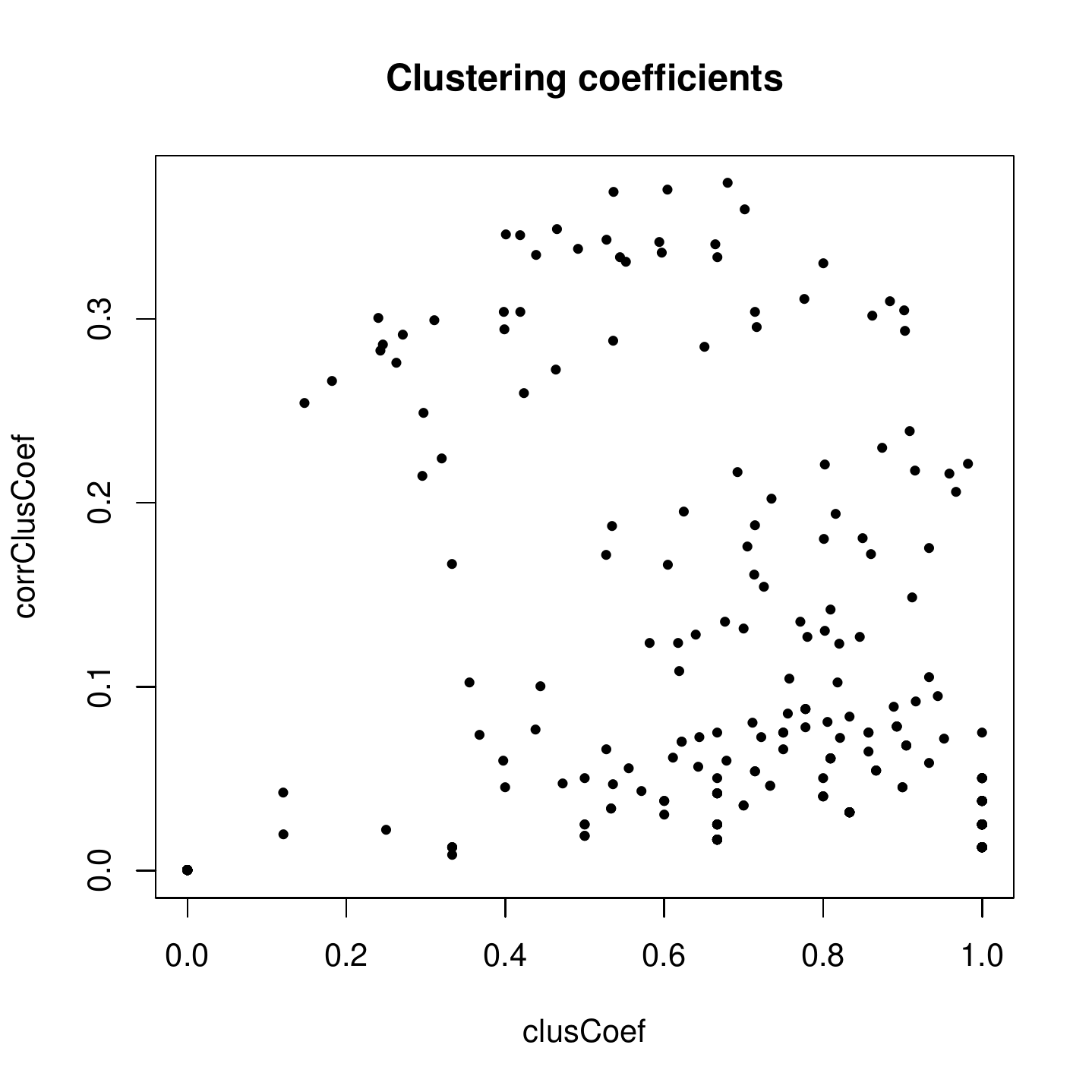}\qquad\includegraphics[width=0.47\textwidth,viewport=0 5 390 360,clip=]{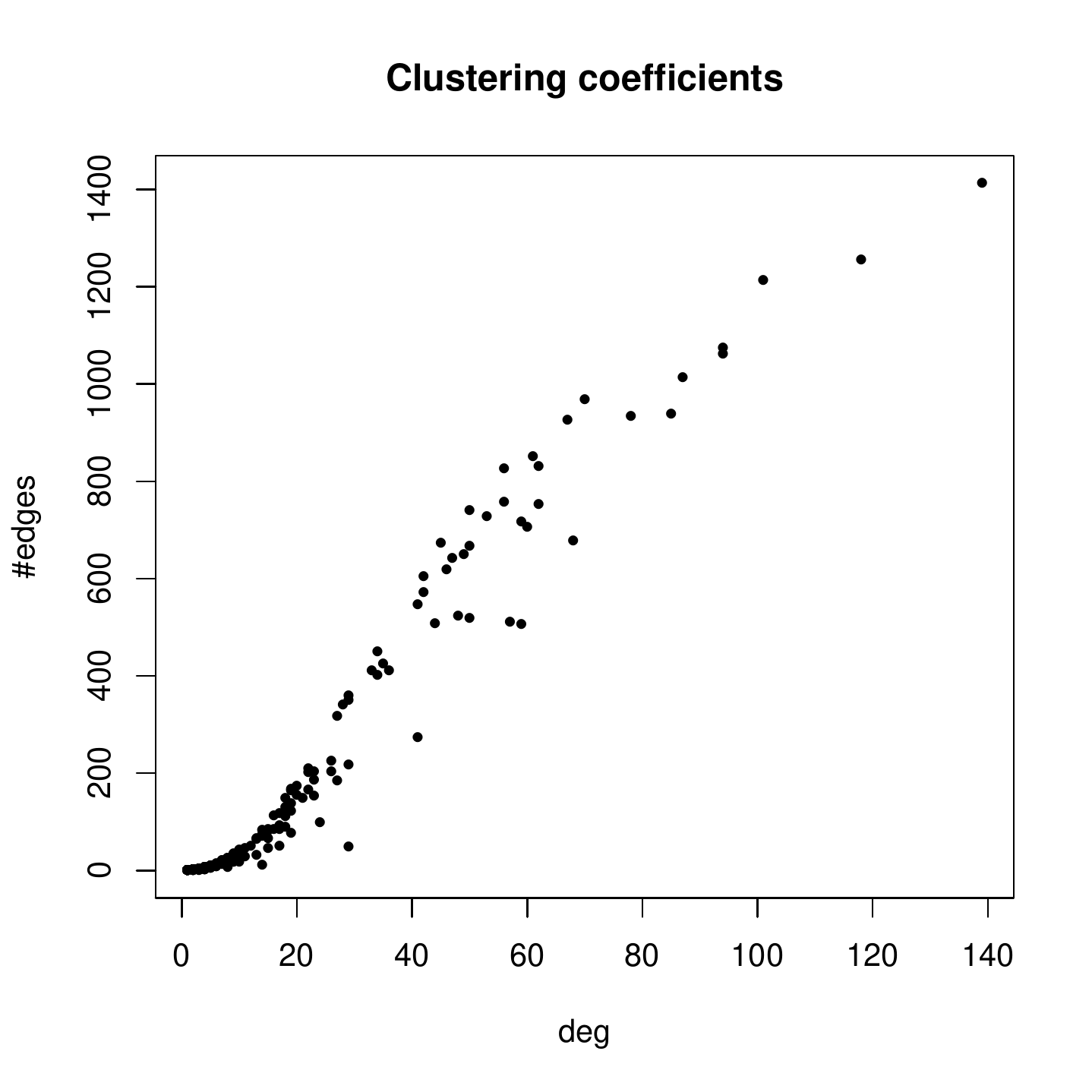}}
\caption{Comparison -- ordinary and corrected clustering coefficients; degrees and number of edges \label{ccc}}
\end{figure}

In Figure~\ref{ccc} the set $\{ (cc(e), cc'(e)) : e \in \Edges \}$  is displayed for the US Airports 1997 network.
The correlation between both coefficients is very small. An important observation is that edges with the largest value
of the clustering coefficient have relatively small values of the corrected clustering coefficient.
We also see that the number of edges in a node's neighborhood  is almost functionally dependent on its degree.

\begin{figure}
\centerline{\includegraphics[width=0.47\textwidth,viewport=0 5 390 360,clip=]{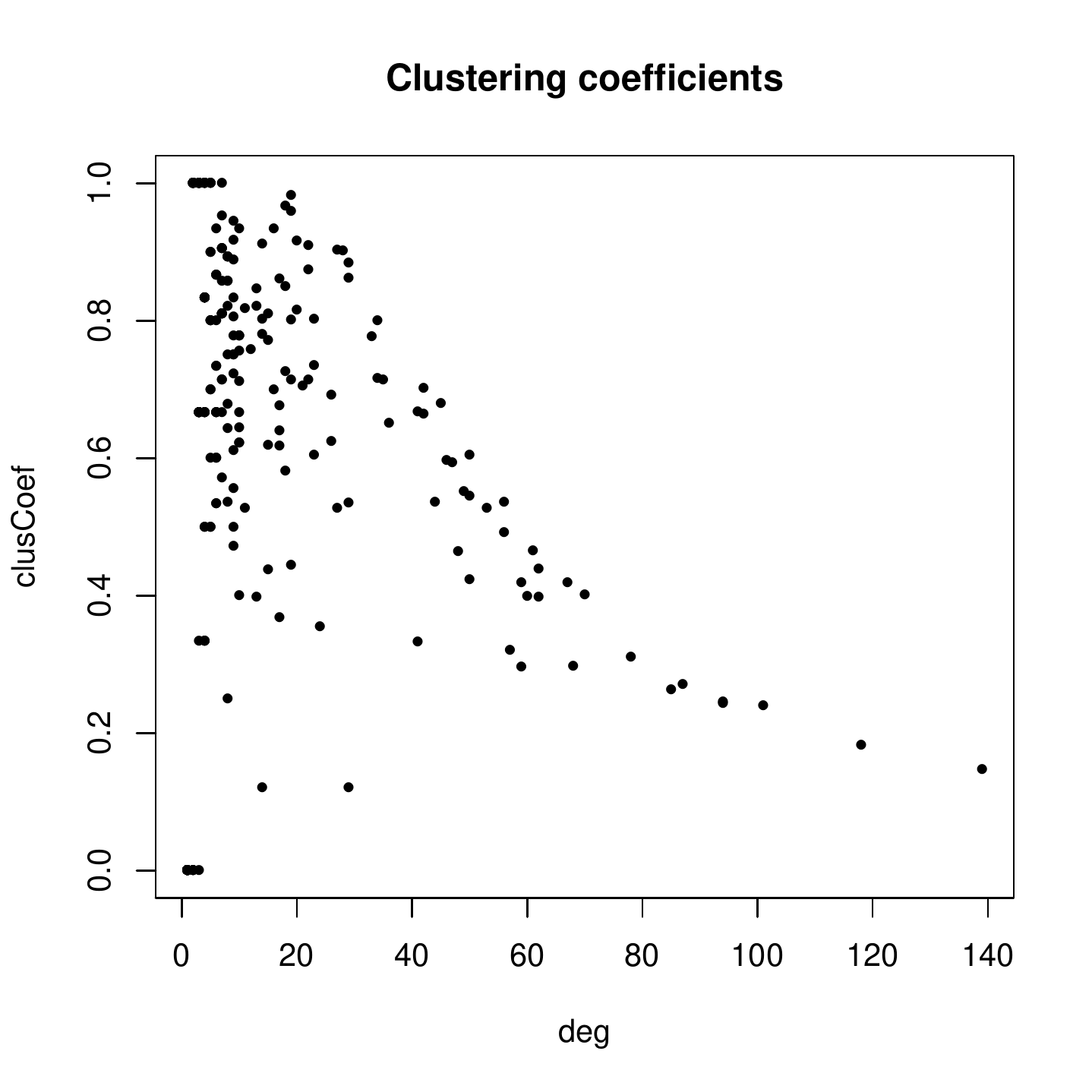}\qquad\includegraphics[width=0.47\textwidth,viewport=0 5 390 360,clip=]{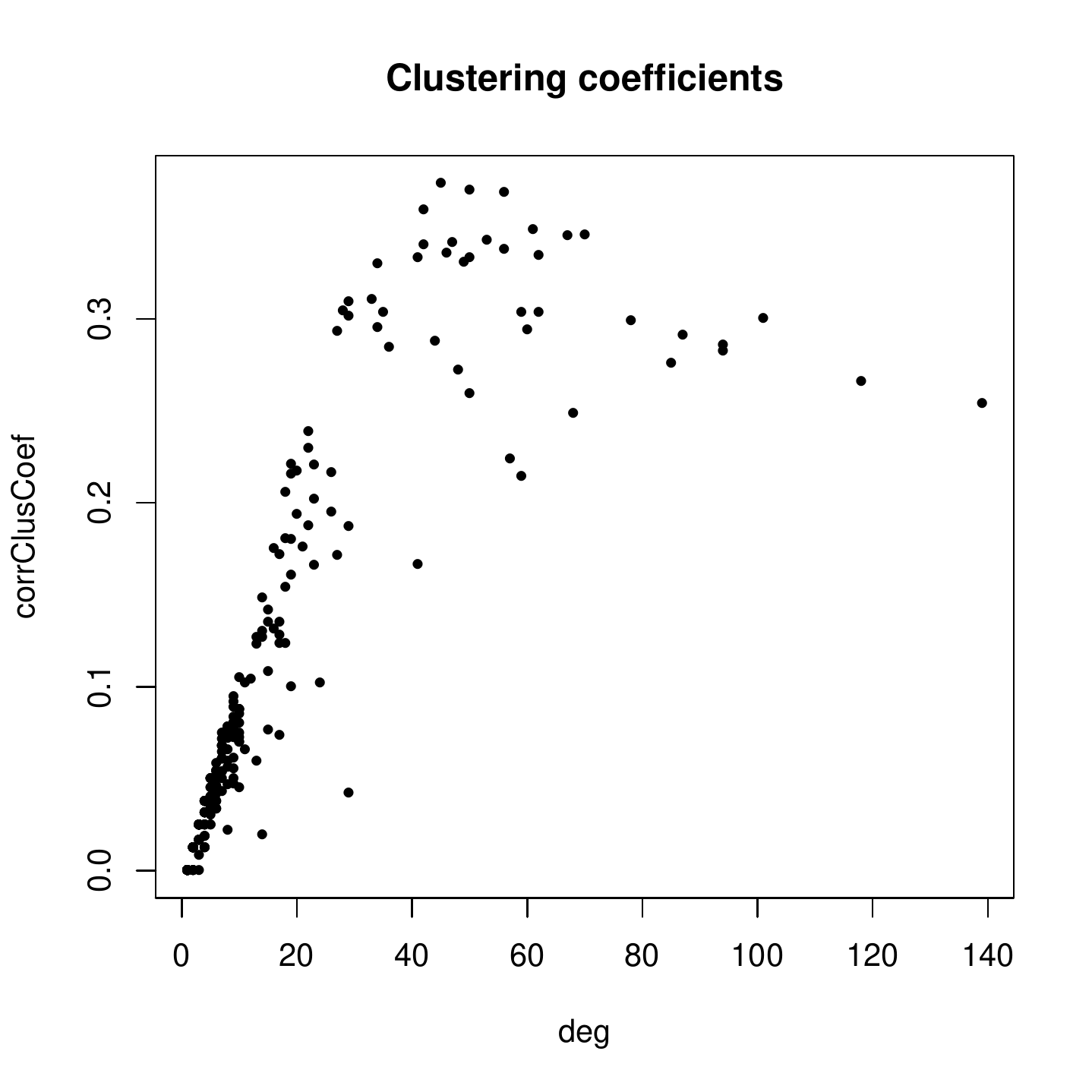}}
\caption{Comparison -- degrees\label{ccd}}
\end{figure}

From Figure~\ref{ccd} we see that the clustering coefficient is decreasing with the increasing degree. Nodes with large
degree have small values of clustering coefficient. The values of corrected clustering coefficient are large for nodes of large degree.


\section{Conclusions}

In the paper we showed that two network measures, the overlap weight and clustering coefficient, are not
suitable for the data analytic task of determining important elements in a given network. We proposed corrected versions
of these two measures that give expected results.

Because $\mu \leq \Delta$ we can replace
in the corrected measures  $\mu$ with $\Delta$. Its advantage is that it can be easier computed; but the
corresponding corrected index  is less `sensitive'.

An interesting task for future research is a comparision of the proposed measures with measures from graph drawing \citep{edgemet,untangle,adaptive}.\medskip


\ifblind\else
\subsection*{Acknowledgments}

The computations were done combining Pajek \citep{ESNA3} with short programs in Python and R \citep{git}.

This work is supported in part by the Slovenian Research Agency (research program
P1-0294 
and research projects
J1-9187,  %
and J7-8279) %
and by Russian Academic Excellence Project '5-100'.

The paper is a detailed and extended version of the talk presented at  the CMStatistics (ERCIM) 2015 Conference.
The author's attendance on the conference was partially supported by the
COST Action IC1408 -- CRoNoS.
\fi

\end{document}